\documentclass[
reprint,
showkeys,
aps,
prd,
longbibliography,
nofootinbib,
onecolumn
]{revtex4-2}

\usepackage[T1]{fontenc}
\usepackage{xparse}
\usepackage{amsmath}
\usepackage{amssymb}
\usepackage{comment}
\usepackage{xstring}

\usepackage{slashed}
\usepackage{cancel}
\usepackage{tensor}

\usepackage{dsfont}
\usepackage{upgreek}
\usepackage{bm}
\usepackage{mathtools}

\usepackage[hyperref,dvipsnames]{xcolor}
\definecolor{goethe-blau}{cmyk}{1.0,0.2,0.0,0.4}
\definecolor{hellgrau}{cmyk}{0.04,0.04,0.05,0.02}
\definecolor{sandgrau}{cmyk}{0.12,0.09,0.13,0.0}
\definecolor{dunkelgrau}{cmyk}{0.25,0.25,0.30,0.75}
\definecolor{emo-rot}{cmyk}{0.04,1.0,0.8,0.07}
\definecolor{purple}{cmyk}{0.08,1.0,0.3,0.36}
\definecolor{senfgelb}{cmyk}{0.01,0.25,1.0,0.05}
\definecolor{gruen}{cmyk}{0.62,0.4,0.87,0.09}
\definecolor{magenta}{cmyk}{0.08,0.86,0.12,0.12}
\definecolor{orange}{cmyk}{0.0,0.7,1.0,0.04}
\definecolor{sonnengelb}{cmyk}{0.0,0.12,0.95,0.0}
\definecolor{helles-gruen}{cmyk}{0.4,0.17,0.81,0.07}
\definecolor{lichtblau}{cmyk}{0.8,0.0,0.06,0.04}

\usepackage[
colorlinks,
pdfpagelabels,
breaklinks,
pdfstartview=FitH,
bookmarksopen=true,
bookmarksnumbered=true,
bookmarksopenlevel=2,
plainpages=false,
hypertexnames=false,
citecolor=emo-rot,
linkcolor=goethe-blau,
urlcolor=purple,
pdftitle={Inhomogeneous condensation in the Gross-Neveu model in noninteger spatial dimensions 1<= d <3}
,pdfauthor={Laurin Pannullo},
pdfkeywords={Gross-Neveu model, stability analysis, inhomogeneous phases, mean-field, noninteger spatial dimensions}
]{hyperref}

\usepackage{bookmark}
\usepackage[capitalise]{cleveref}
\usepackage{multirow}
\usepackage{xstring}
\usepackage{graphicx}	
\usepackage{dcolumn}	

\newcommand{\vecb}{\mathbf}

\newcommand{\x}[1]{\vecb{#1}}
\newcommand{\abs}[1]{\lvert#1\rvert}

\newcommand{\Abs}[1]{\left\lvert#1\right\rvert}
\newcommand{\ifc}{\text{if }}
\newcommand{\otherwisec}{\text{otherwise }}
\newcommand{\iu}{\mathrm{i}}

\newcommand{\N}{N}
\newcommand{\Nf}{N_f}
\newcommand{\Nc}{N_c}
\newcommand{\Ng}{N_{\gamma}}

\newcommand{\action}{\mathcal{S}}
\newcommand{\actionBosonized}{\action_{\phi}}

\NewDocumentCommand{\twopid}{ o } {%
	\IfNoValueTF {#1} {%
		2\uppi%
	}
	{%
		(2\uppi)^{#1}%
	}%
}
\newcommand{\intMeasureOverPi}[2]{\tfrac{\dr[#2][#1]}{\twopid[#1]}\,}

\newcommand{\Heaviside}[1]{\Theta\left(#1\right)}

\newcounter{numrefs}
\makeatletter
\newcommand{\Rcite}[1]{%
	\setcounter{numrefs}{0}
	\@for\@temp:=#1\do{\stepcounter{numrefs}}
	\ifnum\value{numrefs}>1%
	Refs.~\cite{#1}%
	\else%
	Ref.~\cite{#1}%
	\fi%
}
\makeatother

\newcommand{\mstar}{\bar{\sigma} }
\newcommand{\mstarMin}{\bar{\Sigma} }
\newcommand{\mstarZero}{\bar{\sigma}_0 }

\newcommand{\actionEffective}{\action_\mathrm{eff}}

\newcommand{\homo}{\bar}
\newcommand{\homueff}{\homo{U}_\text{eff}}

\newcommand{\Ell}{L}
\newcommand{\ellZero}[1][ ]{l_{0#1}}
\newcommand{\ellOne}[1][ ]{l_{1#1}}
\newcommand{\ellTwo}[1][ ]{l_{2#1}}
\newcommand{\ellZeroArgs}[2]{\ellZero{}\!\left(#1,#2\right)}
\newcommand{\ellTwoArgs}[4]{\ellTwo{}\!\left(#1,#2,#3,#4\right)}
\newcommand{\ellOneArgs}[3]{\ellOne{}\!\left(#1,#2,#3\right)}
\newcommand{\EllTwo}{\Ell_{2}}
\newcommand{\EllTwoArgs}[4]{\EllTwo{}(#1,#2,#3,#4)}

\newcommand{\gtwo}{\ensuremath{\Gamma^{(2)}}}
\newcommand{\gtwoArgs}[4]{\ensuremath{\Gamma^{(2)}(#1,#2,#3,#4)}}

\newcommand{\muc}{\mu_c}
\newcommand{\Gammaf}[1]{\Gamma\left(#1\right)}
\newcommand{\hypgeoWOArgs}{%
	{\vphantom{F}}_{2}\kern-\scriptspace F_{1}%
}
\newcommand{\hypgeo}[4]{%
	\hypgeoWOArgs\left(#1,#2;#3;#4\right)%
}

\newcommand{\Sd}[1]{S_#1}

\makeatletter
\NewDocumentCommand{\dr}{ o o } {%
	\mathop{}\!\mathrm{d}%
	\IfNoValueTF {#2} {%
	}
	{%
		^{#2}\!%
	}
	\IfNoValueTF {#1} {%
	}
	{%
		#1\,%
	}%
}

\NewDocumentCommand{\Dr}{ o o } {%
	\mathop{}\!\mathcal{D}%
	\IfNoValueTF {#2} {%
	}
	{%
		^{#2}\!%
	}
	\IfNoValueTF {#1} {%
	}
	{%
		#1\,%
	}%
}
\makeatother
\newcommand{\BetafunctionWOArgs}{%
	B%
}

\NewDocumentCommand{\Betafunction}{ o m m} {%
	\IfNoValueTF {#1} {%
		\BetafunctionWOArgs\left(#2, #3\right)%
	}{%
		\BetafunctionWOArgs\left(#1; #2, #3\right)%
	}%
}

\DeclareMathOperator{\const}{const}

\DeclareMathOperator{\Det}{\mathrm{Det}}
\providecommand{\ordersymbol}{\mathcal{O}}
\providecommand{\iu}{\mathrm{i}}

\renewcommand{\actionBosonized}{\action_{\sigma}}
\newcommand{\Nbar}{\Ng}

\renewcommand{\Nf}{N}

\usepackage[acronym]{glossaries-extra}
\glssetcategoryattribute{acronym}{nohyperfirst}{true}
\setabbreviationstyle[acronym]{long-short}

\newacronym{qcd}{QCD}{Quantum Chromodynamics}
\newacronym{frg}{FRG}{Functional Renormalization Group}
\newacronym{dse}{DSE}{Dyson-Schwinger equation}
\newacronym{cp}{CP}{critical point}
\newacronym{lp}{LP}{Lifshitz point}
\newacronym{njl}{NJL}{Nambu-Jona-Lasinio}
\newacronym{gn}{GN}{Gross-Neveu}
\newacronym{hbp}{HBP}{homogeneously broken phase}
\newacronym{ip}{IP}{inhomogeneous phase}
\newacronym{sp}{SP}{symmetric phase}
\newacronym{ff}{FF}{Four-Fermion}

\begin{document}

	\title{
		Inhomogeneous condensation in the Gross-Neveu model in noninteger spatial dimensions $1\leq d <3$
	}
	
	\author{Laurin Pannullo}
	\email{pannullo@itp.uni-frankfurt.de}
	
	\affiliation{
		Institut für Theoretische Physik, Goethe-Universität Frankfurt am Main,
		\\
		Max-von-Laue-Straße 1, D-60438 Frankfurt am Main, Germany.
	}
	
	\date{\today}
	
	\begin{abstract}
		The Gross-Neveu model in the $N \to \infty$ limit in $d=1$ spatial dimensions exhibits a chiral inhomogeneous phase (IP), where the chiral condensate has a spatial dependence that spontaneously breaks translational invariance and the $\mathbb{Z}_2$ chiral symmetry.
		This phase is absent in $d=2$, while in $d=3$ its existence and extent strongly depends on the regularization and the value of the finite regulator.
		This work connects these three results smoothly by extending the analysis to noninteger spatial dimensions $1 \leq d <3$, where the model is fully renormalizable.
		To this end, we adapt the stability analysis, which probes the stability of the homogeneous ground state under inhomogeneous perturbations, to noninteger spatial dimensions.
		We find that the IP is present for all $d<2$ and vanishes exactly at $d=2$.
		Moreover, we find no instability towards an IP for $2\leq d<3$, which suggests that the IP in $d=3$ is solely generated by the presence of a regulator.
	\end{abstract} 
	\keywords{Gross-Neveu model, inhomogeneous phases, moat regime, stability analysis,  noninteger spatial dimensions, mean-field 
	}

	\maketitle
	
	\section{Introduction}
	
	A chiral \gls{ip} features a condensate with a spatial dependence that spontaneously breaks translational invariance in addition to chiral symmetry (see \Rcite{Buballa:2014tba} for an extensive review).
	While phases with inhomogeneous order parameters are well established in condensed matter physics, they are a rather exotic phenomenon in high-energy contexts.
	In \gls{qcd} an \gls{ip} occurs in the limit of infinite number of colors $\Nc$ and at asymptotically large chemical potential \cite{Deryagin:1992rw}.
	For the physical case of $\Nc=3$, there are also indications for the realization of such a phase at low temperature and high baryon chemical potential as provided by a \gls{dse} based study that used specific ansatz functions for the inhomogeneous condensate \cite{Muller:2013tya}.
	Recent technical developments  \cite{Motta:2023pks} might even enable an ansatz-free investigation of the \gls{ip} within the \gls{dse} framework.
	Moreover, a functional renormalization group study of \gls{qcd} \cite{Fu:2019hdw} found a so-called moat regime, where the wave-function renormalization assumes negative values.
	Such a regime is closely related to the existence of an \gls{ip} and the implications of such a non-trivial dispersion relation might also be measurable in an experiment \cite{Pisarski:2020gkx,Pisarski:2021qof,Rennecke:2021ovl,Rennecke:2023xhc}. 
	Furthermore, it was shown that  inhomogeneous ground states can naturally be found in theories with $\mathcal{PT}$-type symmetries, which is also realized in finite-density QCD \cite{Schindler:2019ugo,Schindler:2021otf}.

	However, due to the lack of first principle calculations of \gls{qcd} at low temperature and high chemical potential, it is not clear whether the \gls{ip} is indeed realized in nature or what the extent of the moat regime might be.
	Therefore, more often \glspl{ip} are investigated in \gls{ff} and related Yukawa-models, some of which can be regarded as toy-models for \gls{qcd} \cite{Buballa:2014tba}.
	A prominent example is the $(1+1)$-dimensional \gls{gn} model \cite{Gross:1974jv} in the infinite $\Nf$ limit (equivalent to a mean-field approximation in this model), where all quantum fluctuations of the bosonic degrees of freedom are neglected.
	It features a \gls{hbp} at low temperature and baryon chemical potential, where the constant, nonzero chiral condensate breaks the discrete $\mathbb{Z}_2$ chiral symmetry that is realized in the model.
	This phase is separated from a chirally \gls{sp} by a second order line at high temperatures and low chemical potentials that bends down to lower temperatures for increasing chemical potential and ends in a \gls{cp} \cite{Wolff:1985av,Thies:2003kk}.
	If the chiral condensate is restricted to being homogeneous, a first order phase transition extends from this \gls{cp} down to zero temperature.
	However, for spatially dependent condensates, the \gls{cp} coincides with a \gls{lp} from which an \gls{ip} opens up to lower temperatures and higher chemical potentials \cite{Thies:2003kk,Schnetz:2004vr,Thies:2006ti}.
	The coincidence of these points is a feature of the \gls{gn} model (and \gls{njl}-type models) in various dimensions \cite{Thies:2006ti,Koenigstein:2021llr,Nickel:2008ng,Nickel:2009wj,Adhikari:2016jzc} that can be broken up by introducing additional vector interactions \cite{Carignano:2010ac,Carignano:2018hvn}.
	These points can also separate as the result of artifacts at finite regulators in certain regularization schemes \cite{Nickel:2008ng,Adhikari:2016jzc,Buballa:2020nsi}.
	In addition, the model also exhibits a moat regime within a region in the phase diagram that is larger than the \gls{ip} itself \cite{Koenigstein:2021llr}.
	While the phase diagram of the $(1+1)$-dimensional \gls{gn} and the related chiral \gls{gn} model (sometimes also called NJL$_2$ model), which features a continuous chiral symmetry, is fairly understood in the infinite-$\Nf$ limit, it is under intense investigation for finite $\Nf$. Currently, there is no final consensus about which phases persist with full bosonic quantum fluctuations \cite{Karsch:1986hm,Lenz:2020bxk,Lenz:2020cuv,Stoll:2021ori,Ciccone:2022zkg,Lenz:2021kzo,Nonaka:2021pwm,koenigsteinNonperturbativeAspectsLowdimensional2023}.
	However, recent work \cite{koenigsteinNonperturbativeAspectsLowdimensional2023} showed that the feature of negative wave-function renormalization and moat regimes at large $\mu$ is robust under the influence of bosonic fluctuations.
	
	In contrast to the infinite $\Nf$ results in $1+1$ dimensions stands the phase diagram of the same \gls{gn} model in $2+1$ dimensions, where no \gls{ip} for any chemical potential and nonzero temperature is present \cite{Urlichs:2007zz,Narayanan:2020uqt,Buballa:2020nsi,Pannullo:2023one}.
	One only finds a second order line separating the \gls{hbp} at low temperature and chemical potential from the \gls{sp}, which ends in a \gls{cp} at zero temperature \cite{Klimenko:1987gi,Rosenstein:1988dj}.
	It was found that keeping regulators such as the lattice spacing or the Pauli-Villars mass at a finite value, causes the \gls{cp} to be located at a nonzero temperature and the emergence of an \gls{ip} \cite{Buballa:2020nsi}.
	An extended analysis in $2+1$ dimensions revealed that a large class of \gls{ff} models featuring Lorentz-(pseudo)scalar interactions and their Yukawa model extensions do not exhibit an \gls{ip} \cite{Pannullo:2023one}.
	Thus, the absence of an \gls{ip} in the $(2+1)$-dimensional \gls{gn} model is apparently part of a more general behavior of \gls{ff} models in $2+1$ dimensions.
	Still it is not clear what the cause of the absence of the \gls{ip} compared to $1+1$ dimensions is. 
	There has also been considerable effort in understanding the phase structure of the $(2+1)$-dimensional \gls{gn} model beyond the infinite-$\Nf$ limit for finite temperature, chemical potential and magnetic field with lattice and functional methods (see e.g.~\cite{Scherer:2012fjq,Lenz:2023gsq,Hands:2003dh,Strouthos:2003js,Lenz:2023wvk,Kogut:1999um,Hands:1992ck,Hands:1998he}).
	However, there is no concrete evidence for inhomogeneous condensation for finite $\Nf$.
	
	In $3+1$ dimensions, the \gls{gn} and \gls{njl} model exhibit an identical phase diagram in the chiral limit within the mean-field approximation \cite{Nickel:2009wj}.
	In general, one finds a similar phase structure as for the \gls{gn} model in $d=1$ with all three phases and a \gls{cp} present.
	These models are, however, non-renormalizable in $d=3$ and thus one has to keep the employed regulator (e.g.\ the Pauli-Villars mass) at a finite value.
	The phase structure of the theory is strongly dependent on the chosen regularization scheme and value of the regulator \cite{Partyka:2008sv,Kohyama:2015hix,Pannullo:2022eqh,Pasqualotto:2023hho}.
	Varying these can lead to a disappearance of the \gls{cp} for the homogeneous phase transition \cite{Kohyama:2015hix}, a splitting of \gls{lp} and \gls{cp} \cite{Nickel:2008ng,Nickel:2009wj,Carignano:2014jla,Buballa:2020nsi}, and an absence of the \gls{ip} altogether \cite{Partyka:2008sv,Pannullo:2022eqh}.
	
	In this work, we connect these three results from integer dimensions and illustrate why the model shows these qualitatively different phase diagrams.
	To this end, we consider the \gls{gn} model in the mean-field approximation in noninteger number of spatial dimensions $1\leq d <3$.
	This builds on the results of \Rcite{Inagaki:1994ec} where the dependence of the homogeneous phase diagram on $d$ was investigated.
	We extend this by an investigation of the \gls{ip} and the moat regime based on the bosonic two-point function.
	
	The so-called stability analysis, which probes the stability of a homogeneous field configuration against spatially inhomogeneous perturbations by inspection of the bosonic two-point functions, is a common technique to study \glspl{ip}. 
	This method was already used to investigate the \gls{ip} in integer spatial dimensions $d=1,2,3$ within the \gls{gn} and related models (see, e.g., \Rcite{Koenigstein:2021llr,Buballa:2020nsi,Pannullo:2021edr,Pannullo:2022eqh,Buballa:2018hux,Heinz:2015lua,deForcrand:2006zz,Wagner:2007he,Buballa:2020xaa}) and we extend this technique to noninteger spatial dimensions $1\leq d <3$.
	The model is renormalizable for $1\leq d <3$ and the analysis can be formulated independently from details like the fermion representation. 
	Thus, in this setup the only parameter left is the number of spatial dimensions, which allows us to study its influence isolated from other effects.
	At this point it needs to be noted that the concept of noninteger spatial dimensions is something peculiar -- especially since we are investigating a spatial phenomenon.
	Therefore, we should consider the number of spatial dimensions $d$ merely as a parameter that we can vary to interpolate between the physically relevant integer dimensions.
	The study is restricted to zero temperature as it suffices to demonstrate the central findings and makes it possible to give closed form expression for most of the derived quantities.
	
	We find that the instability towards the \gls{ip} gradually disappears when going from $d=1$ to $d=2$.
	Since this setup depends only on $d$ as a parameter, we can identify the number of spatial dimensions as the sole cause of the disappearance of the \gls{ip} in $d=2$.
	Furthermore, there is no instability for $2<d<3$, which suggests that the presence of an \gls{ip} in studies of $(3+1)$-dimensional models is caused by the presence of finite regulators.
	
	This paper is structured as follows.
	\Cref{sec:GN} introduces the \gls{gn} model in $d$ spatial dimensions.
	The homogeneous effective potential at zero temperature and aspects of the homogeneous phase transition are discussed in \cref{sec:ueff} .
	The key quantities of the stability analysis are introduced in \cref{sec:stabanalysis} and the main results of the stability analysis are presented in \cref{sec:results}, which is split between spatial dimensions $1 \leq d \leq 2$ and $2 \leq d < 3$.
	\cref{sec:conclusion} provides a brief conclusion and outlook on future extensions to this work.
	The \cref{App:Ueff,app:gamma2} present technical aspects of the derivation of the effective potential, the stability analysis and the wave-function renormalization.
	
	\section{The Gross-Neveu model in $1 \leq d < 3$ spatial dimensions}
	\label{sec:GN}
	We consider the action of the \gls{gn} model in $D=d+1$ spacetime dimensions
	\begin{equation}
		\action[\bar \psi , \psi] = \int_0^\beta \dr \tau \int \dr^d x \Bigg[ \bar \psi (\slashed \partial + \gamma_0 \mu) \psi 	- \frac{\lambda}{2 \Nf}\left( \bar \psi \psi \right)^2 \Bigg],
	\end{equation}
	where $\psi$ are fermionic spinors with $\Nf \times \Ng$ degrees of freedom (number of flavors\footnote{Note that within the \gls{gn} model, ``flavors'' is the traditional name for this degree of freedom in which the interactions are diagonal. Hence, these flavors are distinctively different from an isospin degree of freedom or quark flavors in \gls{qcd}.} $\times$ dimension of the representation of the Clifford algebra).
	The Euclidean time direction, i.e., the zeroth direction, is compactified with its extent $\beta$ corresponding to the inverse temperature $\beta=1/T$ and
	the $d$-dimensional spatial integration goes over the $d$-dimensional volume $V$.
	In the actual calculations, we will assume both $V$ and $\beta$ to be infinite and hence consider the theory at zero temperature in an infinite volume.
	A baryon chemical potential $\mu$ is introduced in the standard way and the coupling $\lambda$ controls the strength of the \gls{ff} interaction.
	
	By applying a Hubbard-Stratonovich transformation, we remove the \gls{ff} interaction and introduce a real, scalar bosonic field $\sigma$ in the action
	\begin{equation}
		\actionBosonized[\bar \psi , \psi, \sigma] 	=\int_0^\beta \dr \tau \int \dr^d x \left[ \frac{\Nf}{2 \lambda} \sigma^2+\bar \psi (\slashed \partial + \gamma_0 \mu + \sigma ) \psi \right] ,
	\end{equation}	
	where the introduced bosonic field fulfills the Ward identity
	\begin{equation}
		\left\langle \bar \psi (x) \psi (x) \right\rangle = \frac{-\Nf}{\lambda} \left\langle \sigma(x) \right\rangle
	\end{equation}
	that connects the expectation values of the chiral condensate and the bosonic field at the spacetime point $x$.
	The model possesses a discrete $\mathbb{Z}_2$ chiral symmetry in integer dimensions under the transformation
	\begin{align}
		\psi \to \gamma_5 \psi\,,\quad \bar \psi \to -\bar \psi \gamma_5\,,\quad \sigma \to -\sigma,
	\end{align}
	where $\gamma_5$ is the Dirac matrix that anti-commutes with the spacetime Dirac matrices.
	Thus, the auxiliary field $\sigma$ also serves as an order parameter of the spontaneous breaking of the chiral symmetry.
	The special connection between chirality and the number of spacetime dimensions, as well as the ambiguities of defining $\gamma_5$ \cite{tHooft:1972tcz} in noninteger dimensions cause the chiral symmetry to be strictly present only in integer dimensions.
	Nevertheless, in analogy to this symmetry, we denote phases with $\langle \sigma \rangle \neq 0$ as \gls{hbp} (or \gls{ip}, if $\langle \sigma \rangle$ is spatially dependent) as well as phases with $\langle \sigma \rangle=0$ as \gls{sp} even in noninteger dimensions.
	
	Moreover, one has to choose a reducible representation of the Clifford algebra in odd spacetime dimensions in order to find an additional matrix that anti-commutes with the spacetime Dirac matrices.
	This is particularly relevant in $2+1$ dimensions, where one needs to use a reducible $4\times4$ representation to regain the notion of chirality \cite{Buballa:2020nsi,Pannullo:2021edr,Pisarski:1984dj,Gies:2009da}.
	Even though our analysis will be independent of specific representations and their dimensions, we will assume a representation that enables the existence of a matrix $\gamma_5$ in the respective integer dimensions.
	Irrespective of the number of dimensions and representation, we can assume the standard anti-commutation relation for the spacetime Dirac matrices $\{\gamma_\mu,\gamma_\nu\}=2 \delta_{\mu\nu} \mathds{1}$ to hold \cite{tHooft:1972tcz}.
	
	Integrating over the fermionic fields in the path integral yields the so-called effective action 
	\begin{equation}
		\frac{\actionEffective[\sigma]}{\Nf} =\int_0^\beta \dr \tau \int \dr^d x \, \frac{\sigma^2}{2 \lambda} - \ln \Det \left[\beta \left(\slashed \partial + \gamma_0 \mu + \sigma \right)\right], \label{eq:effectiveAction}
	\end{equation}
	where $\Det$ denotes a functional determinant.
	In the following, we consider only the leading term in a $1/\Nf$ expansion (equivalent to a mean-field approximation in this case), which neglects all quantum fluctuations of $\sigma$. 
	Then, the only field configurations $\Sigma$ that contribute to the path integral are these that minimize the effective action $\actionEffective$ globally.
	In the case of a broken symmetry, there are multiple such field configurations which are connected by the transformations of the broken symmetry.
	One typically picks one of these configurations in the evaluation of observables (compare, e.g., \Rcite{Nickel:2009wj,Koenigstein:2021llr}).
	This is equivalent to introducing an explicit breaking to the action and extrapolating this term to zero.
	
	The model is renormalizable for $d<3$ \cite{Hands:1991py} and we use as a renormalization condition that the vacuum expectation value of the auxiliary field assumes a finite homogeneous value $\langle \sigma \rangle |_{T=\mu=0}=\mstarZero$.
	The UV-divergent contributions from loop integrals are regularized with a spatial momentum cutoff.
	This regularization scheme is chosen due to its simplicity and its application being independent of the number of spatial dimensions.
	The scheme restricts the spatial loop momenta to a $d$-dimensional sphere with radius $\Lambda$ in the regularized integrals and $\Lambda$ is then sent to infinity in the renormalization procedure.

	\subsection{The homogeneous effective potential at zero temperature}
	\label{sec:ueff}
	We define the homogeneous effective potential $\homueff$ as the effective action of the homogeneous bosonic field per volume and degree of freedom, i.e.,
	\begin{align}
		\homueff(\mstar,\mu,d) \coloneq \frac{\mathcal{\actionEffective}\left[\bar \sigma\right]}{\Nf V \beta} ,
	\end{align}
	where $\bar \sigma$ is the bosonic field restricted to homogeneous field configurations, i.e., $\bar \sigma  =\const$.
	We proceed to calculate the homogeneous effective potential at zero temperature in the infinite spatial volume
	\begin{align}
		\homueff(\mstar,\mu,d) ={}&  \frac{\mstar^2}{2 \lambda} - \frac{1}{\beta V}\ln \Det \left(\slashed \partial + \gamma_0 \mu + \mstar \right)=\nonumber\\
		={}&\frac{\mstar^2}{2 \lambda} -\frac{\Nbar}{2} \int \intMeasureOverPi{d}{ p}   \left[E - \Theta(\mu^2-E^2) (E-|\mu|) \right]= \nonumber\\
		={}& \frac{\mstar^2}{2 \lambda} -\frac{\Nbar}{2} \, \ellZeroArgs{\bar\sigma^2}{\mu} \label{eq:ueffUnReg},
	\end{align}
	where $E^2=\mstar^2 + \x{p}^2$.
	The integral $\ellZero$ is obviously UV-divergent for every number of spatial dimensions $d>0$.
	We renormalize the effective potential with the condition $\langle \sigma \rangle |_{T=\mu=0}=\mstarZero$ (see \cref{sec:GN}).
	This condition corresponds to $\min_{\mstar} \homueff\, |_{T=\mu=0} = \mstarMin\,|_{T=\mu=0}  = \mstarZero$ within the infinite $\Nf$ limit.
	Therefore, $\mstarZero$ fulfills the homogeneous gap equation
	\begin{align}
		\frac{\dr \homueff }{\dr \mstar} \Bigg|_{T=\mu=0,\mstar=\mstarZero} ={}&\left[\frac{\mstar }{\lambda} - \mstar\Nbar \int_{-\infty}^\infty\intMeasureOverPi{}{p_0} \int_\Lambda \intMeasureOverPi{d}{p} \frac{1}{(p_0 - \iu \mu)^2 + E^2} \right] \Bigg|_{T=\mu=0,\mstar=\mstarZero} = \nonumber \\
		={}& \left[ \mstar \left(\frac{ 1 }{\lambda} - \Nbar \ellOne\right) \right] \Bigg|_{T=\mu=0,\mstar=\mstarZero} \stackrel{!}{=} 0 \label{eq:gapequation},
	\end{align}	
	which is used to tune the coupling $\lambda$ in order to renormalize the theory.
	
	\Cref{App:Ueff} outlines the calculation of $\ellZero$ and $\ellOne$ for spatial dimensions $1\leq d<3$, which are needed to obtain the renormalized effective potential
	\begin{align}
		\begin{split}
			\homueff(\mstar,\mu,d) 
			= \frac{\Ng}{ 2^d \uppi ^{\frac{d}{2}}}  \Biggr[&\frac{(d+1) \Gamma \left(-\frac{d+1}{2}\right)  }{ 8 \sqrt{\uppi }} \left( - \frac{\mstarZero^{d-1} \mstar^2}{2} +  \frac{|\mstar|^{d+1}}{d+1}  \right)  + \\
			& +  \frac{\Theta\left(\bar{\mu}^2\right)  }{d   \Gamma \left(\frac{d}{2}\right)}\, |\mstar|^{d+1} \left|\frac{\bar{\mu}}{\mstar}\right|^{d}\left( \hypgeo{-\tfrac{1}{2}}{\tfrac{d}{2}}{\tfrac{d+2}{2}}{-\tfrac{\bar{\mu}^2}{\mstar^2}}  -\Abs{\frac{\mu}{\mstar}}\right)\Biggl],
		\end{split}\label{eq:ueffarbitraryd}
	\end{align}
	where $\hypgeoWOArgs$ is the Gaussian hypergeometric Function defined by \cref{eq:hypgeo}, $\bar{\mu}^2=\mu^2-\mstar^2$ and a divergent, thermodynamically irrelevant constant term is neglected.
	The effective potential in noninteger spatial dimensions was first investigated in \Rcite{Inagaki:1994ec}.
	However, a closed form expression for $T=0$ and finite chemical potential was not explicitly given and thus we provide it for completeness.
	
\begin{figure}
		\centering
		\includegraphics[width=3.4in,trim=1cm 1.0cm 1cm 1.5cm, clip]{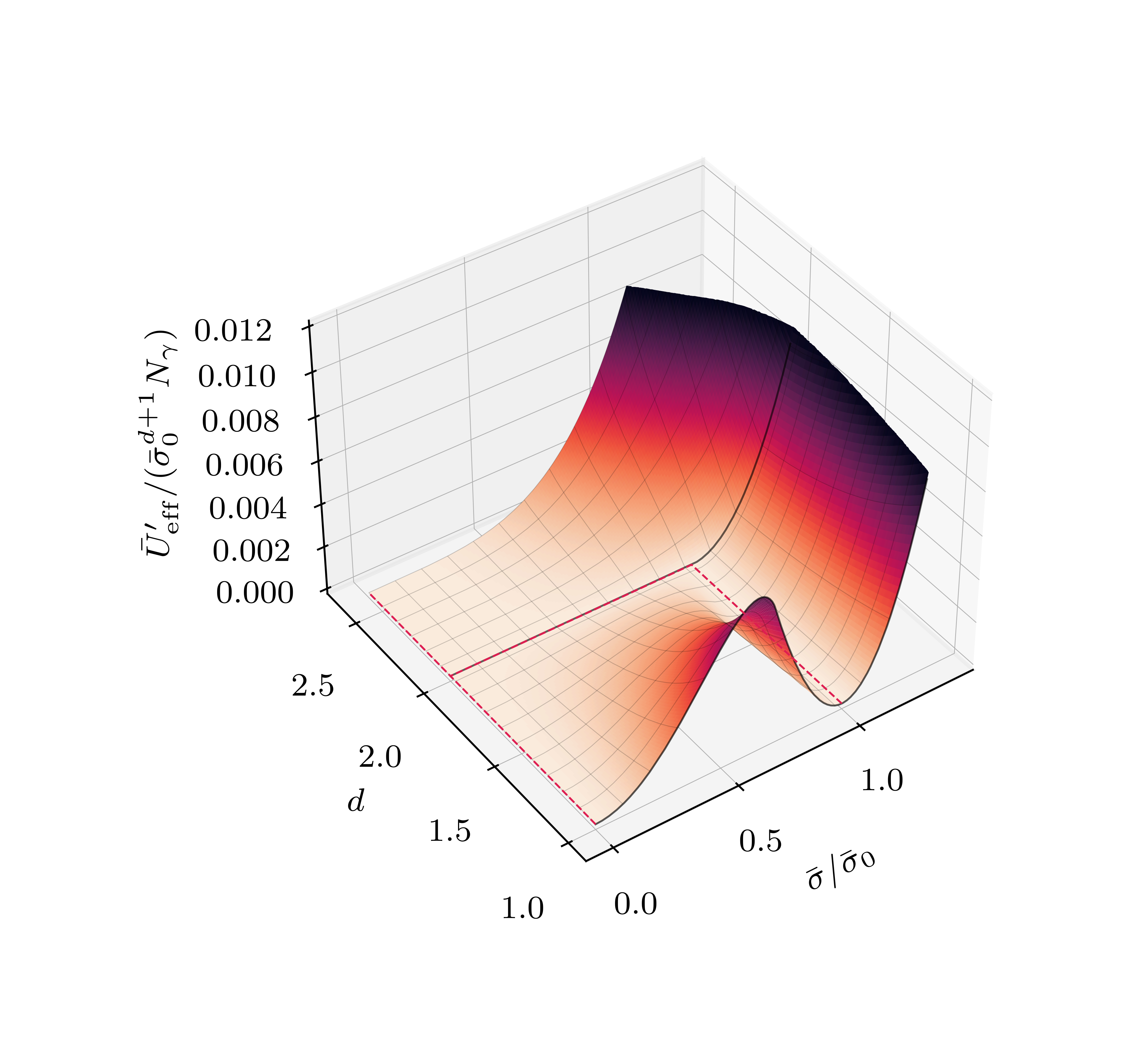}
		\caption{The renormalized effective potential $\homueff'(\mstar,\muc(d),d)= \homueff(\mstar,\muc(d),d)-\homueff(0,\muc(d),d)$ in the $\mstar,d$-plane at the critical chemical potential $\muc(d)$, where the homogeneous phase transition occurs.
			The red dashed lines indicate the minima $\mstarMin(d)$. }
		\label{fig:ueff}
	\end{figure}

	For homogeneous fields, one finds by inspection of $\homueff$ for all number of spatial dimensions $1 \leq d < 3$ an \gls{hbp} at low chemical potential indicated by the minimizing field value $\mstarMin$ being nonzero.
	For chemical potentials larger than a critical chemical potential $\muc(d)$, the system enters the \gls{sp} signaled by $\Abs{\mstarMin}=0$ (see \Rcite{Inagaki:1994ec} for a detailed discussion of the homogeneous phase structure).
	\Cref{fig:ueff} shows the renormalized effective potential $\homueff'(\mstar,\muc(d),d)= \homueff(\mstar,\muc(d),d)-\homueff(0,\muc(d),d)\,$\footnote{The symmetric contribution is subtracted  in order to facilitate the comparison between different $d$.} in the $\mstar,d$-plane at the critical chemical potential $\muc(d)$ with the red dashed lines indicating the minima $\mstarMin(d)$.
	This illustrates how the phase transition is of first order for $d<2$ due to the potential barrier separating the two minima at $\mstarMin=0,\mstarZero$.
	The potential is flat for $|\mstar|\leq\mstarZero$ at $d=2$, which is caused by a combined effect of zero temperature and the \gls{cp} being located at this point.
	\Rcite{Inagaki:1994ec} documents how this \gls{cp} evolves from $(\mu,T)/\mstarZero\approx(0.608, 0.318)$ in $d=1$ to $(\mu,T)/\mstarZero=(1.0,0)$ in $d=2$.
	For $d>2$ the \gls{cp} vanishes and the homogeneous phase transition is strictly of second order.
	
	\subsection{Stability analysis at zero temperature }
	\label{sec:stabanalysis}
	The key concept of the stability analysis is to apply an arbitrary inhomogeneous perturbation of infinitesimal amplitude to a homogeneous field configuration $\mstar$ and analyze the curvature of the effective action $\actionEffective$ under this perturbation.
	If the global homogeneous minimum $\mstarMin$ is used as an expansion point, a negative curvature indicates that there exists an inhomogeneous field configuration with an even lower action and thus confirms the existence of an \gls{ip}.
	For a detailed derivation of the stability analysis in the \gls{gn} model in $1+1$ and $2+1$ dimensions we refer to \Rcite{Buballa:2020nsi,Koenigstein:2021llr}. 
	Here, we present only the final result for the bosonic two-point function $\gtwo$, which is the previously mentioned curvature of the effective action in the direction of an inhomogeneous perturbation of momentum $\x{q}$ to the homogeneous bosonic field $\mstar$.
	One finds that this curvature is only dependent on the magnitude of the bosonic momentum $|\x{q}|= q$ and not its direction in the $d$-dimensional space.
	This circumstance makes it possible to apply this technique in noninteger spatial dimensions.
	
	The two-point function at zero temperature has the general form
	\begin{equation}
		\gtwoArgs{\mstar^2}{\mu}{q^2}{d} = \frac{1}{\lambda} - \Nbar \ellOneArgs{\mstar^2}{\mu}{d} + \EllTwoArgs{\mstar^2}{\mu}{q^2}{d}, \label{eq:generalGtwo}
	\end{equation}
	where we recognize the same contribution $1/\lambda - \Ng l_1$ as in the gap equation and that the whole momentum dependence resides in $\EllTwo$, which is given by
	\begin{align}
		\EllTwoArgs{\mstar^2}{\mu}{q^2}{d} = \tfrac{1}{2}\left(q^2 + 4 \mstar^2\right)\Nbar \int_{-\infty}^{\infty} \intMeasureOverPi{}{p_0} \int \intMeasureOverPi{d}{ p} \frac{1}{((p_0-\iu \mu)^2 + \mstar^2 + (\x{p} + \x{q})^2) ((p_0-\iu \mu)^2 + \mstar^2 + \x{p}^2) } .\label{eq:L2_first}
	\end{align}
	The evaluation of this expression for arbitrary $1\leq d<3$ is outlined in \cref{app:gamma2}, while for the integer cases of $d=1$ we refer to \Rcite{Koenigstein:2021llr} and for $d=2$ to \Rcite{Buballa:2020nsi}.
	We find for arbitrary spatial dimensions $1\leq d<3$ that the two-point function evaluates to
	\begin{align}
		\gtwoArgs{\mstar^2}{\mu}{q^2}{d} = \frac{\Nbar}{2^d \uppi ^{\frac{d}{2} }\Gamma \left(\frac{d}{2}\right) }  &\left[\frac{\Gamma \left(\frac{1-d}{2}\right) \Gamma \left(\frac{d+2}{2}\right) }{d\uppi  } \left(|\mstarZero|^{d-1}-| \mstar| ^{d-1}\right) + \right. \label{eq:gamma2arbitraryd}\\
		& + 
		\begin{rcases}\begin{cases}
				\dfrac{| \mu| ^{d-1}}{(d-1)} \ & \text{if } \mstar=0 ,\, \mu\neq 0\\
				\dfrac{|\mstar|^{d-1} }{d }\,\left|\dfrac{\bar{\mu}}{\mstar}\right|^{d}  \hypgeo{\frac{1}{2}}{\frac{d}{2}}{\frac{d+2}{2}}{-\frac{\bar{\mu}^2}{\mstar^2}}\ & \text{if } \mstar\neq0 ,\, \bar{\mu}^2 > 0\\
				0 \ & \text{otherwise }
		\end{cases} \end{rcases} + \nonumber\\
		& \left.+ \left(\tfrac{q^2}{4} +  \mstar^2 \right)\int_{0}^{1} \dr[x] \times \begin{rcases}\begin{cases}
				\dfrac{\tilde{\mu}^{d-3}}{(3-d)} \hypgeo{\frac{3}{2}}{\frac{3-d}{2}}{\frac{3-d}{2}+1}{-\frac{\tilde{\Delta}^2}{\tilde{\mu}^2}} -\dfrac{\tilde{\mu}^{d-2}}{\abs{\mu}}   &\ifc \tilde{\mu}^2> 0 \\
				\dfrac{\tilde{\Delta}^{d-3}}{2} \Betafunction{\frac{d}{2}}{\frac{3-d}{2}}  &\otherwisec 
		\end{cases}\end{rcases}\right], \nonumber
	\end{align}
	where  $\tilde{\Delta}^2=\mstar^2 +  q^2 x (1-x)$ and $\tilde{\mu}^2=\mu^2-\tilde{\Delta}^2$.
	The remaining integral over $x$ is evaluated numerically since no closed form can be given for the integral.
	
	The other quantity of interest is the bosonic wave-function renormalization $z$, where negative values indicate a moat regime.
	This $z$ is the coefficient of the bosonic kinetic term $\propto 1/2 \,\partial_\mu \sigma \partial_\mu \sigma$ in the effective action that is contained in the fermionic contribution\footnote{This term can be explicitly seen in an expansion of the $\ln \Det$ contribution in the effective action \labelcref{eq:effectiveAction}, e.g., in a Ginzburg-Landau approach. See also \Rcite{koenigsteinNonperturbativeAspectsLowdimensional2023} for a study of $Z$ in the $(1+1)$-dimensional \gls{gn} model at finite $\N$, i.e., in the presence of bosonic fluctuations.}.
	We can calculate $z$ from the bosonic two-point function by differentiating it twice with respect to $q$ and setting $q=0$ \cite{Koenigstein:2021llr}, i.e.,
	\begin{align}
		z(\mstar,\mu,d)={}&\frac{1}{2}\frac{\dr^2\, \gtwoArgs{\mstar}{\mu}{q^2}{d}}{\dr q^2}\Bigg|_{q=0} = \nonumber\\
		={}& \frac{\Nbar}{2^{d+2} \uppi^{\frac{d}{2}}\, \Gammaf{\frac{d}{2}}}   \times 
		\begin{cases}
			\!\begin{aligned}[b]
				& \frac{1}{(3-d) \bar{\mu}^{3-d}}\, \hypgeo{\tfrac{3}{2}}{\tfrac{3-d}{2}}{\tfrac{5-d}{2}}{-\tfrac{\mstar^2}{\bar{\mu}^2}} + \\[1ex]
				-{}&  \frac{1}{(5-d)}   \frac{\mstar^2}{\bar{\mu}^{2}} \frac{1}{\bar{\mu}^{3-d}} \, \hypgeo{\tfrac{5}{2}}{\tfrac{5-d}{2}}{\tfrac{7-d}{2}}{-\tfrac{\mstar^2}{\bar{\mu}^2}}+ \\[1ex]
				+{}& \frac{\bar{\mu}^{d-2}}{|\mu|}\left[\frac{\mstar^2}{\mu^{2}} \left(1+\frac{1}{3 \bar{\mu}^{2}}  \left(2\mstar^2 - (4-d) \mu^2  \right)\right)-1 \right]
			\end{aligned}           & \ifc \bar \mu ^2>0\\[3ex]
			\dfrac{1}{2 |\mstar|^{3-d}} \left[\Betafunction{\frac{d}{2}}{\frac{3-d}{2}} -  \Betafunction{\frac{d}{2}}{\frac{5-d}{2}}\right]   & \otherwisec
		\end{cases}. \label{eq:z}
	\end{align}
	The derivation of $z$ is outlined in \cref{app:gamma2}.
	If $z$ is evaluated on the global homogeneous minimum $\mstarMin(\mu,d)$, we denote it as $Z(\mu,d)\equiv z\left(\mstarMin(\mu,d),\mu,d\right)$.
	
	\section{Results of the stability analysis}
	\label{sec:results}
	In this section the results that are obtained by the stability analysis of the \gls{gn} model for $1\leq d < 3$ spatial dimensions are discussed.
	The discussion is split in $1\leq d \leq 2$ and $2\leq d < 3$ due to the different conclusions that we can draw from these two intervals.
	
	\subsection{$1\leq d \leq 2$}
\begin{figure}
		\centering
		\begin{minipage}[t]{.48\textwidth}
			\centering
			\vspace{0pt}
			\includegraphics[width=3.4in]{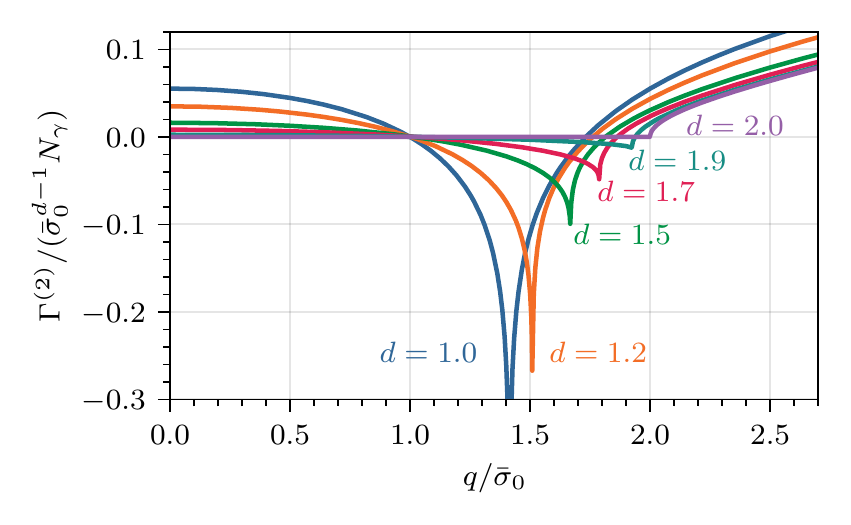}
			\caption{The two-point function $\gtwoArgs{\mstarMin^2=0}{\mu_c^+}{q^2}{d}$ as a function of the bosonic momentum $q$ for various spatial dimensions $1 \leq d \leq 2$ evaluated at the homogeneous minimum $\mstarMin$ at chemical potential $\mu=\mu_c^+$ (the critical chemical potential with an infinitesimal positive shift).
				Compare to Ref.~\cite{Koenigstein:2021llr} for $d=1$ and Refs.~\cite{Buballa:2020nsi,Pannullo:2023one} for $d=2$.}
			\label{fig:gamma2atmucsmalld}
		\end{minipage}%
		\hfill
		\begin{minipage}[t]{.48\textwidth}
			\centering
			\vspace{0pt}
			\includegraphics[width=3.4in]{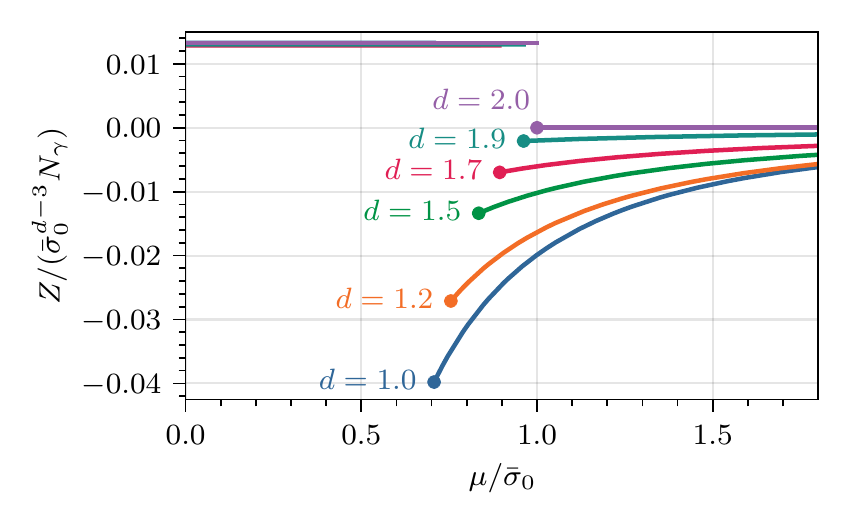}
			\caption{The wave-function renormalization $Z$ as a function of the chemical potential for various spatial dimensions $1 \leq d \leq 2$. The circle indicates the chemical potential $\mu_c$ at which the homogeneous phase transition is located.
				Compare to Ref.~\cite{Koenigstein:2021llr} for $d=1$.}
			\label{fig:zvsmusmalld}
		\end{minipage}
	\end{figure}

	\Cref{fig:gamma2atmucsmalld} shows the two-point functions $\gtwoArgs{\mstarMin^2}{\mu_c^+}{q^2}{d}$ for $1 \leq d \leq 2$ spatial dimensions at $\mu=\mu_c^+$, which is the critical chemical potential with an infinitesimal positive shift.
	This ensures that the homogeneous minimum used as the expansion point is $\mstarMin=0$. 
	For $d=1$, the two-point function diverges logarithmically at $q=2\mu$ for all $\mu>\muc$ as also observed in \Rcite{Koenigstein:2021llr}.
	For $1<d<2$, the integral over $x$ in \cref{eq:gamma2arbitraryd} has to be performed numerically.
	It is thus not immediately clear, whether the two-point function diverges as in $d=1$ for $q=2\mu$.
	The integrand in \cref{eq:gamma2arbitraryd} is divergent at $x=1/2$ for $\mstar=0,q=2\mu$ and expanding it at $x_0=1/2$ reveals that the most divergent term is $\propto \left(x-1/2\right)^{d-2}$.
	Hence, the integral over $x$ is finite for any $d>1$.
	Thus, the divergence of the two-point functions vanishes for $d>1$, but a cusp that is a negative minimum is retained.
	This preserves the instability at $\mu_c$ for $1<d<2$.
	However, one finds that the offset of $\gtwo$ at $q=0$ increases with increasing $\mu$.
	Thus, for $1<d<2$ there is an upper $\mu$ at which the \gls{ip} vanishes (see \cref{fig:pd}).
	
	It was documented in \Rcite{Koenigstein:2021llr} that in the $(1+1)$-dimensional \gls{gn} model there is a region of the \gls{ip} in the $\mu$-$T$-plane that is not detected by the stability analysis.
	This is where the homogeneous minimum $\mstarMin$ assumes a finite value, which is separated from the true inhomogeneous minimum by an energy barrier.
	Thus, $\mstarMin$  appears to be stable against inhomogeneous perturbations even though an inhomogeneous condensate is energetically preferred \cite{Thies:2003kk}.
	We expect this to happen in some portion of the phase diagram for all spatial dimensions $1<d<2$, since the first order phase transition, which was identified as the cause of this effect in $d=1$, is also present there.
	
	By increasing $d$ further to $d=2$, the two-point function evolves to the known $(2+1)$-dimensional result \cite{Buballa:2020nsi,Pannullo:2023one}.
	The two-point function is constant and zero for all $q\leq2\mu$ at which it rises for $q>2\mu$. 
	This corresponds to a degeneracy of the homogeneous minimum and field configurations with small inhomogeneous perturbations of momentum $q\leq2\mu$.
	Hence, it cannot provide any information about the energetically preferred state.
	However, it was found that the crystal kink solutions of the $(1+1)$-dimensional \gls{gn} model embedded in $2$ spatial dimensions are energetically degenerate to homogeneous field configurations even for finite amplitudes at $(\mu,T)=(\mu_c,0)$ in the $(2+1)$-dimensional \gls{gn} model \cite{Urlichs:2007zz}.\footnote{It might be interesting to embed the solutions of the $(1+1)$-dimensional \gls{gn} model in $1<d<2$ similar to what was done for $d=2$ in \cite{Urlichs:2007zz}.
		In this way, one could observe how the degeneracy between homogeneous configurations and these inhomogeneous modulations develops at $d=2$.} 
	This observation and our results for the two-point function suggest a flat effective potential for a variety of inhomogeneous modulations.
	This would be similar to the flat homogeneous potential shown in \cref{fig:ueff}, which is caused by the special nature of the \gls{cp} at this point.
	
	We observe that all numbers of spatial dimensions $1 \leq d < 2$, where the \gls{cp} is also located at a nonzero temperature (as discussed in \cref{sec:ueff}), exhibit an instability.
	This is due to the coincidence of the \gls{cp} and the \gls{lp} for the renormalized \gls{gn} model.

	\Cref{fig:zvsmusmalld} shows the wave-function renormalization evaluated at $\mstarMin$ as a function of $\mu$ for $1\leq d\leq 2$.
	We observe $Z<0$ for $\mu>\mu_c$ and $d<2$, which is the key property of a moat regime \cite{Pisarski:2020gkx,Pisarski:2021qof,Rennecke:2021ovl,Rennecke:2023xhc}.
	Thus, a moat regime is retained for all chemical potentials for $d<2$.

	\subsection{$2\leq d < 3$}
\begin{figure}[t]
		\centering
		\begin{minipage}[t]{.48\textwidth}
			\centering
			\vspace{0pt}
			\includegraphics[width=3.4in]{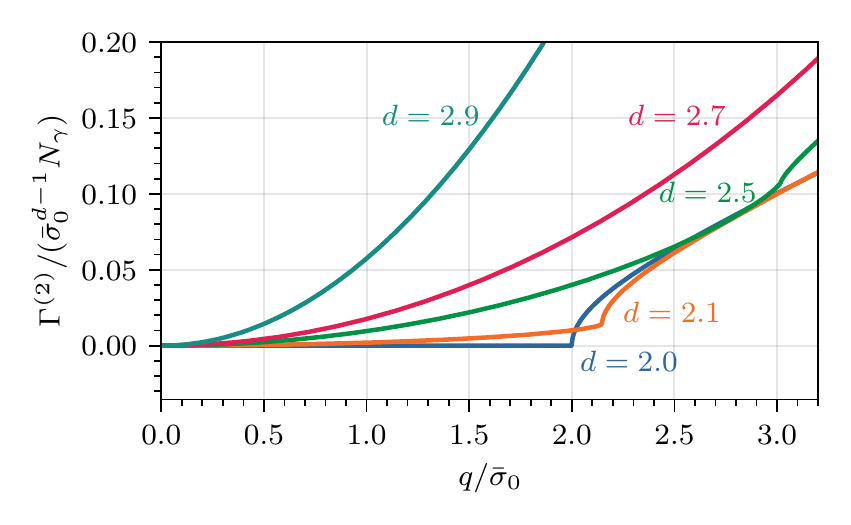}
			\caption{The two-point function $\gtwoArgs{\mstarMin^2=0}{\mu_c^+}{q^2}{d}$ as a function of the bosonic momentum $q$ for various spatial dimensions $2 \leq d < 3$ evaluated at the homogeneous minimum $\mstarMin$ at chemical potential $\mu=\mu_c^+$ (the critical chemical potential with an infinitesimal positive shift).
				Compare to Refs.~\cite{Buballa:2020nsi,Pannullo:2023one} for $d=2$.}
			\label{fig:gamma2atmuclarged}
		\end{minipage}%
		\hfill
		\begin{minipage}[t]{.48\textwidth}
			\centering
			\vspace{0pt}
			\includegraphics[width=3.4in]{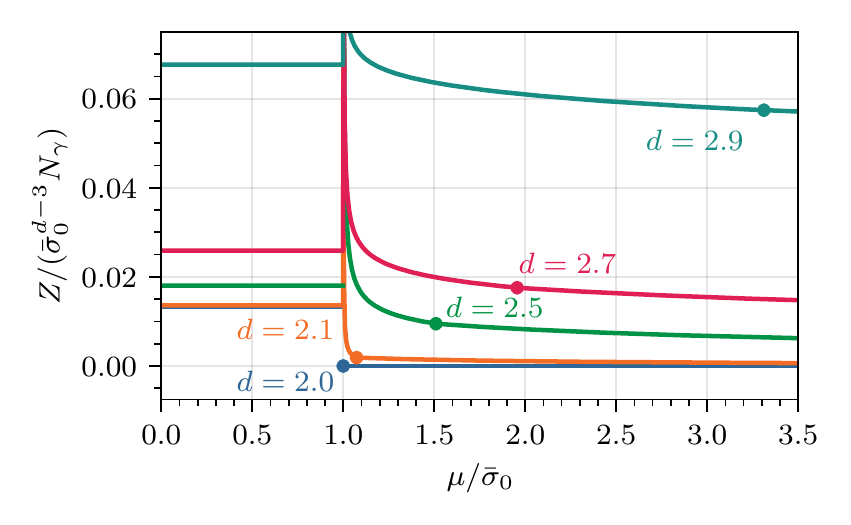}
			\caption{The wave-function renormalization $Z$ as a function of the chemical potential for various spatial dimensions $2 \leq d < 3$. The circle indicates the chemical potential $\mu_c$ at which the homogeneous phase transition is located.}
			\label{fig:zvsmularged}
		\end{minipage}
	\end{figure}

	\Cref{fig:gamma2atmuclarged} shows $\gtwoArgs{\mstarMin^2}{\mu}{q^2}{d}$ for spatial dimensions $2 \leq d < 3$ at $\mu=\mu_c^+$.
	For spatial dimensions $d>2$, the constant behavior vanishes and the two-point function approaches a parabolic shape, but the cusp at $q=2\mu$ remains a non-analytic point.
	Thus, by inspection of the two-point function we recognize that there is no instability for $2<d<3$.
	This is in stark contrast to existing results of the \gls{njl} model in $3+1$ dimensions, which exhibits the same phase diagram as the $(3+1)$-dimensional \gls{gn} model within the mean-field approximation \cite{Inagaki:1994ec,Nickel:2009wj}.
	Here one finds instabilities towards an \gls{ip} \cite{Nakano:2004cd,Pannullo:2022eqh,Buballa:2018hux} and even the energetically preferred inhomogeneous condensates by minimizing the effective action with a suitable ansatz \cite{Sadzikowski:2000ap,Nakano:2004cd,Heinz:2015lua,Partyka:2008sv,Nickel:2009wj,Carignano:2011gr,Carignano:2012sx,Lakaschus:2020caq}.
	Due to the smooth evolution of the two-point function for $1 \leq d <3$, we would not expect a significant change in behavior caused by increasing the number of dimensions when going from $d<3$ to $d=3$. 
	The difference, however, is that while we investigated the renormalized model in $d<3$, it loses its renormalizability in $d=3$. 
	Thus, the aforementioned investigations are performed at a finite regulator. 
	It was shown that varying the regularization scheme (e.g.~Pauli-Villars regularization, spatial momentum cutoff, lattice regularization) and the value of its regulator can have a severe impact on the existence and extent of the \gls{ip} \cite{Pannullo:2022eqh,Partyka:2008sv}.
	Moreover, the \gls{cp} coincides with the \gls{lp} only for some regularization schemes, e.g., Pauli-Villars or dimensional regularization.
	For small enough regulators this \gls{cp} and with it also the \gls{lp} is located at a nonzero temperature, which enables the existence of the \gls{ip}.
	In a similar fashion an investigation of the $(2+1)$-dimensional \gls{gn} model revealed that an \gls{ip} exists at a finite regulator and vanishes when removing the regulator \cite{Buballa:2020nsi,Pannullo:2021edr,Narayanan:2020uqt}. 
	This finding and the lack of instability for $2<d<3$ in the renormalized setup as presented in this work suggest the conclusion that the existence of the \gls{ip} in the $(3+1)$-dimensional \gls{gn} model (and due to their equivalence also the \gls{njl} model) is solely triggered by the choice of the regularization scheme and the presence of a finite regulator.

	\Cref{fig:zvsmularged} depicts the bosonic wave-function renormalization $Z$ as a function of the chemical potential for various spatial dimensions $2\leq d<3$.
	While the minimum value of the wave-function renormalization at $d=2$  is $Z=0$, it is strictly positive for $2<d<3$.
	Therefore, no moat regime is retained for $d>2$.
	Moreover, we note that $Z$ diverges at $\mu/ \mstarZero =1$.
	This is caused in $2<d<3$ by the second order homogeneous phase transition, where the condensate starts to ``melt'' for chemical potentials $\mu/\mstarZero>1$.
	This enables $1=\mu^2\approx\mstarMin(\mu)^2$, which causes divergences in $Z$ (compare to \cref{eq:z}).
	Graphically speaking, this is caused by the cusp that is present in the two-point function at $\Abs{q}=2 \sqrt{\mu^2-\mstar^2}$, being located at $q=0$ for $\mu^2 = \mstarMin^2$.
	Then, this causes $Z$, which is the second derivative of the two-point function, to diverge.

	\section{Conclusion and Outlook}
	\label{sec:conclusion}
	\subsection{Conclusion}
	\glsresetall
	
	Within the stability analysis one applies inhomogeneous perturbations to the homogeneous ground state and inspects the curvature of the effective action for these perturbations.
	If one finds negative values for this curvature, which are given by negatives values in the momentum dependence of the bosonic two-point function, the homogeneous field configuration is unstable and an inhomogeneous ground state is energetically preferred.
	
	We adapted this method to noninteger spatial dimensions and illuminated how the instability towards an \gls{ip} in $1+1$ dimensions turns into an absence of instability in $2+1$ dimensions.
	By continuously increasing the number of spatial dimensions from $d=1$ to $d=2$, we observed how the two-point function evolves as a function of $d$.
	The \gls{ip} is present for all $d<2$ in some range of $\mu$ and the instability vanishes exactly at $d=2$.
	Moreover, for $1<d<2$ there is an upper chemical potential at which the instability vanishes, but a moat regime is retained for all chemical potentials.
	This renormalized setup is independent of regulators and details like the fermion representation, which allows us to study the isolated effect of the number of dimension. 
	Thus, we identified that the sole driver of the disappearance of the \gls{ip} at $d=2$ is the number of spatial dimensions, and by extension the dependence of the \gls{cp} and \gls{lp} on $d$.

	For $2<d<3$, one finds that the two-point function is positive for all bosonic momenta $q\geq0$ and thus there is no instability towards an \gls{ip}. 
	This is qualitatively different from existing results in $d=3$ that exhibit an \gls{ip} \cite{Nickel:2009wj,Nakano:2004cd,Pannullo:2022eqh,Buballa:2018hux,Sadzikowski:2000ap,Nakano:2004cd,Heinz:2015lua,Partyka:2008sv,Nickel:2009wj,Carignano:2011gr,Carignano:2012sx,Lakaschus:2020caq}.
	The difference is the need for a finite regulator in $d=3$ spatial dimensions that can cause the appearance of a \gls{cp} and \gls{lp} at a nonzero temperature.
	This effect was also observed in the \gls{gn} model for $d=2$ where it led to the appearance of the \gls{ip} even though it is not present when the model is renormalized \cite{Narayanan:2020uqt,Buballa:2020nsi,Pannullo:2021edr}.
	This observation and our results suggest that the appearance of the \gls{ip} in the $(3+1)$-dimensional \gls{gn} and \gls{njl} model is generated by the necessary use of a finite regulator.
	
\begin{figure}
		\centering
		\includegraphics[width=3.4in]{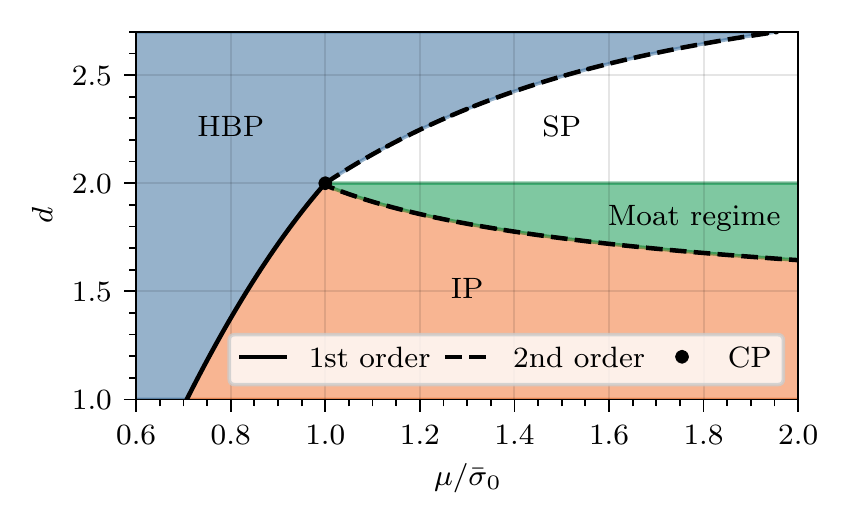}
		\caption{The phase diagram of the renormalized \gls{gn} model as obtained from the stability analysis in the plane of spatial dimensions $d$ and chemical potential $\mu$. The phase diagram shows a homogeneously broken phase (HBP) with $\sigma(x)=\mstar=\const$, a symmetric phase (SP) with $\sigma(x)=0$, an inhomogeneous phase (IP) with $\sigma(x)=f(x)$ and a moat regime with negative wave-function renormalization, i.e., $Z<0$. 
			The boundaries of the HBP are calculated by Eq.~(3.33) and Eq.~(3.35) from Ref.~\cite{Inagaki:1994ec}.}
		\label{fig:pd}
	\end{figure}

	\Cref{fig:pd} summarizes these findings in the phase diagram of the renormalized \gls{gn} model at zero temperature in the plane of the number of spatial dimensions $d$ and chemical potential $\mu$.
	
	Interestingly there is a connection of this work with investigations of the $(3+1)$-dimensional \gls{njl} model that use dimensional regularization to regulate the theory \cite{Inagaki:2007dq,Inagaki:2012re,Kohyama:2015hix,Fujihara:2008ae,Fujihara:2008wx}.
	Due to the nonrenormalizability of the model, the number of spatial dimensions $d$ has to be fixed at a value $d<3$ and additionally one has to introduce a mass scale $M_0$ (because the regulator $d$ itself is dimensionless).
	Both $d$ and $M_0$ are then tuned such that certain observables in the vacuum assume fixed values (e.g.~the pion decay constant).
	In this way, one could interpret the present work as the $(3+1)$-dimensional \gls{gn} model with dimensional regularization, since the applied renormalization also introduced a mass scale $\mstarZero$ and we vary the dimensions $d$.
	In this picture, by analyzing the phase structure for different $d$, we really investigate the regulator dependence of the phase diagram of the $(3+1)$-dimensional \gls{gn} model for the dimensional regularization scheme.
	Vice versa, the effect of considering $(3+1)$-dimensional models with dimensional regularization at finite regulator is that one generates the phase structure of lower dimensional versions of the respective models.
	
	\subsection{Outlook}
	An obvious extension of the present work might be the investigation of the \gls{njl} model.
	It features an additional Yukawa interaction term with a pion field of the form $\bar \psi  \iu \gamma_5 \vec{\tau} \cdot \vec{\pi} \psi$.
	However, the ambiguities of $\gamma_5$ in noninteger dimensions lead to an altered anti-commutation relation $\{\gamma_\mu,\gamma_5\}$ \cite{tHooft:1972tcz}, which significantly changes the renormalization and the stability analysis of the \gls{njl} model compared to integer dimensions.
	While it is still possible to conduct the stability analysis, the results depend on these ambiguities in noninteger dimensions and thus no results for this model are presented.
	A more detailed discussion of the resulting implications can be found in \Rcite{pannulloInhomogeneousChiralCondensates2023}.

	Most investigations of the \gls{ip} in $(3+1)$-dimensional models (see, e.g., \Rcite{Nickel:2008ng,Nickel:2009wj,Carignano:2010ac,Carignano:2011gr,Carignano:2012sx,Carignano:2014jla,Carignano:2018hvn,Buballa:2018hux}) use the Pauli-Villars regularization. 
	In order to connect better to these results, it would be instructive to carry out the present analysis in noninteger spatial dimensions with the Pauli-Villars regularization at a finite regulator.
	In this way, one could show that it is possible to regain an \gls{ip} for $2<d\leq3$ by considering appropriate values of the regulator and smoothly connect our noninteger analysis to established, finite regulator results in $d=3$.
	
	Several investigations in integer dimensions (e.g.~\Rcite{Urlichs:2007zz,Nickel:2009wj}) have embedded $1$-dimensional solutions of the $(1+1)$-dimensional \gls{gn} model in higher dimensional models.
	This procedure is also adaptable to noninteger $d$, since the $(d-1)$-dimensional space perpendicular to the modulation can be treated in a way where $d$ enters only as a parameter just as in the present study.
	This would enable us to observe how the degeneracy of the $1$-dimensional solutions of the $1+1$-dimensional \gls{gn} model and homogeneous field configurations at $(\mu,T)=(\mu_c,0)$ in $d=2$ \cite{Urlichs:2007zz} develops for $1<d<2$.
	
	The extension of the present analysis to nonzero temperature is under way.
	A nonzero temperature will likely not change the conclusion about the (non)existence of the instability, since a nonzero temperature in all known occurrences disfavors an \gls{ip}.
	However, it would reveal how the whole phase diagram evolves between the known results in integer spatial dimensions.

	\begin{acknowledgments}
		I thank Adrian Koenigstein, Marc Winstel and Marc Wagner for their helpful comments on this manuscript and numerous, valuable discussions.
		Furthermore, I acknowledge fruitful, related discussions with Michael Buballa, Zohar Nussinov, Gergely Markó, Mike Ogilvie, Robert Pisarski, Fabian Rennecke, Stella Schindler and David Wagner.
		I acknowledge the support of the \textit{Deutsche Forschungsgemeinschaft} (DFG, German Research Foundation) through the collaborative research center trans-regio  CRC-TR 211 ``Strong-interaction matter under extreme conditions''-- project number 315477589 -- TRR 211.	
		I acknowledge the support of the \textit{Helmholtz Graduate School for Hadron and Ion Research}.	
		I acknowledge the support of the \textit{Giersch Foundation}.
		
	\end{acknowledgments}
	
	\appendix
	
	\section{Derivation of the renormalized, homogeneous effective potential}
	\label{App:Ueff}
	In this Appendix, we outline the derivation of the renormalized, homogeneous effective potential by using a spatial momentum cutoff $\Lambda$ to regularize the theory.
	A more detailed derivation and discussion can be found in \Rcite{pannulloInhomogeneousChiralCondensates2023}.
	The homogeneous effective potential was already investigated in \Rcite{Inagaki:1994ec}, thus it is not original to this work.
	However, we need some of the results in the later derivation and hence it is instructive to also include the full derivation of the renormalized, effective potential here.
	Throughout this Appendix, we make regular use of the integral identities $3.194$ from \Rcite{gradshteynTableIntegralsSeries2007}.
	
	We start the derivation by calculating the integral $\ellZero$ that appears in the expression \cref{eq:ueffUnReg} and find that it evaluates to
	\begin{widetext}
		\begin{align}
			\ellZero={}&\frac{S_d}{(2 \uppi)^d} \int \dr[p] p^{d-1}   \left[E - \Theta(\mu^2-E^2) (E-|\mu|) \right] = \label{eq:Appendix_lzero_calculation} \nonumber\\
			={}& \frac{S_d}{(2 \uppi)^d} \frac{ 1}{d} \left[ \Lambda^d |\mstar| \hypgeo{-\tfrac{1}{2}}{\tfrac{d}{2}}{\tfrac{d+2}{2}}{-\left(\tfrac{\Lambda}{\mstar}\right)^2}  -\Theta\left(\bar{\mu}^2\right) \Abs{\mstar}^{d+1} \Abs{\frac{\bar{\mu}}{\mstar}}^{d}\left( \hypgeo{-\tfrac{1}{2}}{\tfrac{d}{2}}{\tfrac{d+2}{2}}{-\tfrac{\bar{\mu}^2}{\mstar^2}} -\Abs{\frac{\mu}{\mstar}}\right) \right] = \nonumber\\
			\begin{split}
				={}&\frac{S_d}{(2 \uppi)^d} \frac{ 1}{2} \left[  -\frac{ |\mstar|^{d+1} \Gamma \left(-\frac{d}{2}-\frac{1}{2}\right) \Gamma \left(\frac{d}{2}+1\right)}{d \sqrt{\uppi }}
				+ \Lambda^d \left(\frac{2  \Lambda }{d+1}+\frac{ \mstar^2}{(d-1) \Lambda }+\frac{ \mstar^4}{4(3-d) \Lambda ^3}+\ordersymbol\left(\Lambda^{-5}\right)\right) \right.+ \\
				&-\left. \Theta\left(\bar{\mu}^2\right) \Abs{\mstar}^{d+1} \Abs{\frac{\bar{\mu}}{\mstar}}^{d}\left( \hypgeo{-\tfrac{1}{2}}{\tfrac{d}{2}}{\tfrac{d+2}{2}}{-\tfrac{\bar{\mu}^2}{\mstar^2}} -\Abs{\frac{\mu}{\mstar}}\right) \right],
			\end{split}
		\end{align}
		where $S_d={2 \uppi^{\frac{d}{2}}}/{\Gamma \left(\frac{d}{2}\right)}$ is the surface area of a $d$-dimensional unit sphere, $\bar{\mu}^2=\mu^2-\mstar^2$ and we expanded the $\Lambda$ dependent terms for $\abs{\Lambda/\mstar}\gg1$ in the last step.
		$\hypgeoWOArgs$ denotes the Gaussian hypergeometric Function that can be represented via the integral 
		\begin{align}
			\hypgeo{\alpha}{\beta}{\gamma}{z} ={}& \frac{1}{B(\beta,\gamma-\beta)}  \int_{0}^{1} \dr[t] t^{\beta-1} (1-t)^{\gamma-\beta-1} (1-t z)^{-\alpha} \label{eq:hypgeo}
		\end{align}
		with $B$ being the Beta function.

		In order to derive the renormalized, homogeneous effective potential, one needs to tune the coupling $\lambda$ by imposing that the minimum of the renormalized, homogeneous effective potential in vacuum is at $\mstar=\mstarZero$.
		To do so, we employ the gap equation \labelcref{eq:gapequation}, where we need to calculate the integral $\ellOne$.
		For the renormalization procedure, we would only need $\ellOne$ at $\mu=0$ and finite $\mstar$.
		However, we calculate it in its general $\mu$ and $\mstar$ dependent form, since the same integral also appears in the bosonic two-point function that we need for the stability analysis. 
		We find for the integral
		\begin{align}
			\ellOneArgs{\mstar^2}{\mu}{d} ={}&    \int_\Lambda \intMeasureOverPi{d}{p} \int_{-\infty}^{\infty} \intMeasureOverPi{}{p_0} \frac{1}{(p_0 - \iu \mu)^2 +E^2} =   \frac{S_d}{(2 \uppi)^d} \int_0^\Lambda \dr p \, p^{d-1} \frac{1-\Theta\left(\mu^2-E^2\right)}{2 E} = \nonumber \\
			={}&  \frac{S_d}{(2 \uppi)^d} \frac{1}{2d|\mstar|}  \left[\Lambda^d\hypgeo{\tfrac{1}{2}}{\tfrac{d}{2}}{\tfrac{d+2}{2}}{-\left(\tfrac{\Lambda}{\mstar}\right)^2} - \Theta\left(\bar{\mu}^2\right)  \Abs{\bar{\mu}}^d  \hypgeo{\tfrac{1}{2}}{\tfrac{d}{2}}{\tfrac{d+2}{2}}{-\tfrac{\bar{\mu}^2}{\mstar^2}} \right]. \label{eq:l1dd}
		\end{align}
		Using the vacuum part of this result and the gap equation \labelcref{eq:gapequation}, we tune the coupling to the appropriate value
		\begin{align}
			\frac{1}{\lambda} ={}& \Ng \frac{S_d}{(2 \uppi)^d} \frac{1}{2d\mstarZero}  \Lambda^d\hypgeo{\tfrac{1}{2}}{\tfrac{d}{2}}{\tfrac{d+2}{2}}{-\left(\tfrac{\Lambda}{\mstarZero}\right)^2} = \\ ={}& \Ng \frac{S_d}{(2 \uppi)^d} \frac{1}{2} \left[ 
			\frac{ \mstarZero^{d-1} \Gamma \left(\frac{d}{2}+1\right) \Gamma \left(\frac{1}{2}-\frac{d}{2}\right)}{d\sqrt{\uppi }}+
			\Lambda^d\left(\frac{1}{(d-1)  \Lambda }+\frac{ \mstarZero^2}{2(3-d) \Lambda ^3}+\ordersymbol\left({\Lambda^{-5}}\right)\right)
			\right], \label{eq:lambda}
		\end{align}
		where we expanded the $\Lambda$ dependent terms for $\abs{\Lambda/\mstarZero}\gg1$.
		Inserting the expression for $\ellZero$ from \cref{eq:Appendix_lzero_calculation} and the tuned coupling into \cref{eq:ueffUnReg} yields the renormalized, homogeneous effective potential from \cref{eq:ueffarbitraryd}, where a divergent, thermodynamically irrelevant constant term is neglected.
		We find for the symmetric limit $\mstar \to 0$ that the renormalized effective potential is reduced to
		\begin{align}
			\homueff (\mstar=0,\mu,d)
			={}& \frac{\Ng}{ 2^d \uppi ^{\frac{d}{2}}}    \frac{  \Abs{\mu}^{d+1} }{   \Gamma\left(\frac{d}{2}\right)d(d+1) } .
		\end{align}
	\end{widetext}
	
	\section{Derivation of the bosonic two-point function and the bosonic wave-function renormalization}
	\label{app:gamma2}
	In this Appendix, we outline the derivation of the bosonic two-point function and the wave-function renormalization.
	A more detailed derivation and discussion can be found in \Rcite{pannulloInhomogeneousChiralCondensates2023}.
	Throughout this Appendix, we make regular use of the integral identities $3.194$ from \Rcite{gradshteynTableIntegralsSeries2007}.
	
	The bosonic-two point function consists of a constant contribution $1/\lambda-\Ng \ellOne$, which is derived in \cref{App:Ueff}. 
	Thus, we only need to calculate the integral $\EllTwo$ that is given in \cref{eq:L2_first}.
	The first step is to get rid of any contributions that depend on the angle between the loop momentum $\x{p}$ and the external bosonic momentum $\x{q}$.
	We can achieve this by applying a Feynman parametrization of the integral in \cref{eq:L2_first} resulting in
	\begin{widetext}
		\begin{align}
			\ellTwoArgs{\mstar^2}{\mu}{q^2}{d} 
			={}&  \Nbar \int_{-\infty}^{\infty} \intMeasureOverPi{}{p_0} \int \intMeasureOverPi{d}{ p} \int_{0}^{1} \dr[x]\,  \frac{1}{\left[(\x p + \x q)^2 x + \Delta^2 x + (1-x) \x p ^2 + (1-x)\Delta^2\right]^2 }= \nonumber\\
			={}&  \Nbar \int_{-\infty}^{\infty} \intMeasureOverPi{}{p_0} \int \intMeasureOverPi{d}{ p} \int_{0}^{1} \dr[x]\, \frac{1}{\left[p^2 + \Delta^2 + q^2 x (1-x)\right]^2 } \label{eq:l2Feynman}
		\end{align}
		where we performed a shift of the integration variable $\x{p}+\x{q}x \to \x{p}$ and $\Delta^2=(p_0-\iu \mu)^2 + \mstar^2$. 
		In this form we can easily carry out the integration over the temporal momenta and over the spatial momenta subsequently to obtain the form
		\begin{align}
			\ellTwoArgs{\mstar^2}{\mu}{q^2}{d} 
			={}& \Nbar \frac{\Sd{d}}{(2 \uppi)^d} \int_{0}^{1} \dr[x] \int_0^\infty \dr[p] p^{d-1} \frac{1}{4 \tilde{E}^3}\left[\Heaviside{\frac{\tilde{E}}{|\mu|}-1}-\frac{\tilde{E}}{|\mu| }\delta\left(\frac{\tilde{E}}{|\mu|}-1\right)\right] = \nonumber\\
			={}&  \frac{\Nbar }{2^{d+1}  \uppi^{d/2} \Gamma\left(\frac{d}{2}\right)} 
			\int_{0}^{1} \dr[x] \times \begin{cases}
				\dfrac{\tilde{\mu}^{d-3}}{(3-d)}\, \hypgeo{\frac{3}{2}}{\frac{3-d}{2}}{\frac{3-d}{2}+1}{-\frac{\tilde{\Delta}^2}{\tilde{\mu}^2}} -\dfrac{\tilde{\mu}^{d-2}}{\abs{\mu}}   &\ifc \tilde{\mu}^2> 0 \\
				\dfrac{\tilde{\Delta}^{d-3}}{2} \,\Betafunction{\frac{d}{2}}{\frac{3-d}{2}}  &\otherwisec 
			\end{cases}, \label{eq:l2dd}
		\end{align}
		where $\tilde{E}^2=\mstar^2 + p^2 +  q^2 x (1-x)$,  $\tilde{\Delta}^2=\mstar^2 +  q^2 x (1-x)$ and $\tilde{\mu}^2=\mu^2-\tilde{\Delta}^2$.
		Since only certain limits of $\mstar^2$, $\mu^2$ and $q^2$ allow to give a closed form expression of the integral over $x$, we simply evaluate the integral over $x$ numerically.
		Inserting the result for $\ellTwo$ \labelcref{eq:l2dd} and for $1/\lambda -\ellOne$ from \cref{eq:l1dd,eq:lambda} into \cref{eq:generalGtwo} yields the full two-point function from \cref{eq:gamma2arbitraryd}.
		The integral over $x$ is trivial in the limit of $q\to0$ for which we obtain for the two-point function the closed form
		\begin{align}
			\gtwoArgs{\mstar^2}{\mu}{q^2=0}{d} = \frac{\Nbar}{2^d \uppi ^{\frac{d}{2} }\Gamma \left(\frac{d}{2}\right) }  &\left[\frac{\Gamma \left(\frac{1-d}{2}\right) \Gamma \left(\frac{d+2}{2}\right) }{d\uppi  } \left(|\mstarZero|^{d-1}-| \mstar| ^{d-1}\right) + \right. \\
			& + 
			\begin{rcases}\begin{cases}
					\dfrac{| \mu| ^{d-1}}{(d-1)} \ & \text{if } \mstar=0,\, \mu\neq 0\\
					\dfrac{| \mstar| ^{d-1} }{d }\,\Abs{\dfrac{\bar{\mu}}{\mstar}}^d \hypgeo{\frac{1}{2}}{\frac{d}{2}}{\frac{d+2}{2}}{-\frac{\bar{\mu}^2}{\mstar^2}}\ & \text{if } \mstar\neq0 ,\, \bar{\mu}^2 > 0\\
					0 \ & \text{otherwise }
			\end{cases} \end{rcases} + \nonumber\\
			& \left.+ \abs{\mstar}^{d-1} \times  \begin{rcases}\begin{cases}
					\dfrac{1}{(3-d)}\Abs{\dfrac{\bar{\mu}}{\abs{\mstar}}}^{d-3} \hypgeo{\frac{3}{2}}{\frac{3-d}{2}}{\frac{3-d}{2}+1}{-\frac{\mstar^2}{\bar{\mu}^2}} -\dfrac{\bar{\mu}^{d-2}}{\abs{\mu}}   &\ifc \bar{\mu}^2> 0 \\
					\frac{1}{2} \Betafunction{\frac{d}{2}}{\frac{3-d}{2}}  &\otherwisec 
			\end{cases} \end{rcases}\right], \nonumber
		\end{align}
		where $\bar{\mu}^2=\mu^2-\mstar^2$.
		
	\end{widetext}
	
	The bosonic wave-function renormalization $z$ is the curvature of the bosonic two-point function evaluated at $q=0$. 
	By differentiating $\EllTwo$ twice with respect to $q$ and evaluating it at $q=0$, we find 
	\begin{widetext}
		\begin{align}
			z=\frac{1}{2}\frac{\dr^2\, \gtwoArgs{\mstar}{\mu}{q^2}{d}}{\dr q^2}\Bigg|_{q=0} 
			={}& \frac{1}{4} \Nbar  \int_{-\infty}^{\infty} \intMeasureOverPi{}{p_0} \frac{S_d}{(2 \uppi)^d}  \int_0^\infty \dr[p] p^{d-1} \left[  \frac{2}{\left(E^2+(p_0-\iu \mu)^2\right)^2}-\frac{8 \mstar^2 }{3\left(E^2+(p_0-\iu \mu)^2\right)^3}  \right]  \nonumber \\
			={}& \frac{1}{4} \Nbar \frac{S_d}{(2 \uppi)^d}     \int_0^\infty \dr[p] p^{d-1}  \left\{\frac{1}{2 E^3} \left[\Theta\left(\frac{E^2}{\mu^2}-1\right) - \frac{E}{|\mu|}\, \delta \left(\frac{E}{|\mu|}-1\right)\right]   \right. + \nonumber\\
			&\left. -\frac{8 \mstar^2 }{3} \frac{3}{16} \left[ \frac{1}{E^5}\Theta(E^2-\mu^2) - \frac{1}{E^4|\mu|} \,\delta \left(\frac{E}{|\mu|}-1\right) + \frac{1}{3E^3\mu^2} \delta' \left(\frac{E}{|\mu|}-1\right)\right] \right\}.
		\end{align}
	\end{widetext}
	Carrying out the remaining integral over $p$ results in the form given in \cref{eq:z}.
	\bibliography{main}
	
\end{document}